\documentclass[preprint,showpacs,preprintnumbers,amsmath,amssymb]{revtex4}

\usepackage{graphicx}
\usepackage{dcolumn}
\usepackage{bm}

\begin{document}

\vspace{5mm}

\newcommand{\goo}{\,\raisebox{-.5ex}{$\stackrel{>}{\scriptstyle\sim}$}\,}
\newcommand{\loo}{\,\raisebox{-.5ex}{$\stackrel{<}{\scriptstyle\sim}$}\,}

\title{Production of hypernuclei in peripheral relativistic 
ion collisions.}

\author{A.S.~Botvina$^{1,2}$, K.K.~Gudima$^{3}$, J.~Pochodzalla$^{2,4}$}

\affiliation{$^1$Institute for Nuclear 
Research, Russian Academy of Sciences, 117312 Moscow, Russia} 
\affiliation{$^2$Helmholtz-Institut Mainz, J.Gutenberg-Universit{\"a}t, 
55099 Mainz, Germany} 
\affiliation{$^3$Institute of Applied Physics, Academy of Sciences of Moldova, 
MD-2028 Kishinev, Moldova} 
\affiliation{$^4$ Institut f{\"u}r Kernphysik and PRISMA Cluster of 
Excellence, J.Gutenberg-Universit{\"a}t Mainz, D-55099 Germany}

\date{\today}

\begin{abstract}

Within a dynamical and statistical approach we study the main regularities 
in production of hypernuclei coming from projectile and target residues 
in relativistic ion collisions. We demonstrate that yields of hypernuclei 
increase considerably above the energy threshold for $\Lambda$ hyperons, 
and there is a saturation for yields of single hypernuclei with increasing 
the beam energy up to few TeV. Production of specific hypernuclei depend 
very much on the isotopic composition of the projectile, and this gives a 
chance to obtain exotic hypernuclei that may be difficult to reach in 
traditional hypernuclear experiments. Possibilities for the detection of such 
hypernuclei with planned and available relativistic ion facilities are 
discussed. 

\end{abstract}


\maketitle

\section{Introduction}

Nuclear reactions induced by energetic ions lead to abundant production 
of strange baryons (hyperons). When hyperons are captured by nuclei, 
hypernuclei are formed whose lifetime are significantly longer than the 
typical reaction times. These hypernuclei are an important tool to study  
the hyperon--nucleon ($YN$) and hyperon--hyperon ($YY$) interactions at 
low energies 
($Y=\Lambda,\Sigma,\Xi,\Omega$), in order to overcome the limited 
experimental possibilities existing in elementary scattering experiments. 
Double- and multi-strange nuclei are especially interesting, because they 
can provide information about the hyperon--hyperon interaction and 
hyper-matter properties at low temperature. Furthermore, 
hypernuclei can help to investigate the structure of conventional 
nuclei too \cite{Bando,japan}, and extend the nuclear chart into 
the strangeness sector \cite{buyuk2013,cgreiner,greiner}. It is also 
known that hyper-matter should be produced at high nuclear densities, 
which are realized in the core of neutron stars \cite{schaffner}. 
Therefore, production of hypernuclei (particular those with extreme 
isospin) in the laboratory is important for many fields of research. 

Typical observables for hypernuclei are ground-state masses, energy levels, 
and decay properties \cite{japan}. 
The theoretical studies are mainly concentrated on calculating the 
structure of nearly cold hypernuclei with baryon density around the nuclear 
saturation density, $\rho_0 \approx 0.15$ fm$^{-3}$. However, 
a quite limited set of reactions was generally used for producing 
hypernuclei: Reactions with the production of few particles, including kaons, 
are quite effective for 
triggering single hypernuclei, and by using kaon beams one can produce 
double hypernuclei. The goal of this paper is to demonstrate that one can 
essentially extend the frame of hypernuclear studies and produce exotic 
hypernuclei if new many-nucleon reactions are involved. 

We should remember that hyperons were discovered in the 1950-s in reactions 
of nuclear multifragmentation induced by cosmic rays \cite{danysz}. 
During the last 20 years of research a great progress was made in 
investigation 
of the multifragmentation reactions, mainly associated with heavy-ion 
collisions (see, e.g., \cite{smm,aladin97,EOS,ogul} and references therein). 
This gives us an opportunity to apply a well known theoretical method 
adopted for description of these reactions for production of hypernuclei too 
\cite{bot-poch,dasgupta}. On the other hand, it was noticed long ago 
that the absorption of hyperons in spectator regions 
after peripheral nuclear collisions is an effective way for producing 
hypernuclei \cite{wakai1,cassing,giessen,botvina2011}.  Corresponding 
experimental evidences have been reported \cite{Nie76,Avr88}. 
Also central collisions of relativistic heavy ions can lead to productions of 
light hypernuclei \cite{Steinheimer}. 
Recent sophisticated experiments 
have confirmed observations of hypernuclei in such reactions, in both 
peripheral \cite{saito-new} and central collisions \cite{star}. 

We want to pay special attention to formation of hypernuclei in spectator 
region of peripheral relativistic ion collisions. Current research 
concerns light hypernuclei produced in reactions with light projectiles 
\cite{saito-new}, which were previously obtained in other reactions. 
There is also a promising opportunity to study production of large and 
exotic hypernuclei coming from reactions with large projectiles and targets 
\cite{botvina2011,bot2012}. 
In particular, multifragmentation decay of excited hyper-spectator matter 
\cite{buyuk2013,bot-poch,dasgupta}, and the Fermi-break-up of excited light 
hyper-spectators \cite{lorente,botvina2012} are perspective mechanisms. 
Below we undertake a systematic investigation of how new and exotic 
hypernuclei can be obtained in future experiments. For this purpose we 
use a hybrid dynamical and statistical approach, which is widely 
accepted as one of the best tool for description of fragmentation and 
multifragmentation reactions.


\section{Formation of hyper-residues}


A mechanism of peripheral relativistic heavy-ion collisions has 
been established in many experimental and theoretical studies. In the 
simplified picture nucleons from overlapping parts of the projectile 
and target (participant zone) interact strongly with themselves 
and with other hadrons produced in primary and secondary collisions. 
Nucleons from non-overlapping parts do not interact intensively, and 
they form residual nuclear systems, which we call spectator residues 
or spectators. In all transport models 
the production of hyperons is associated with nucleon-nucleon collisions, 
e.g.,  p+n$\rightarrow$n+$\Lambda$+K$^{+}$, or collisions of secondary 
mesons with nucleons, e.g., $\pi^{+}$+n$\rightarrow \Lambda$+K$^{+}$. 
Strange particles may be produced in the participant zone, however, 
particles can re-scatter and undergo secondary interactions. As a result 
the produced hyperons populate the whole momentum space around the 
colliding nuclei, including the vicinity of nuclear spectators. 
Such hyperons can be absorbed by the spectators if their kinetic energy 
(in the rest frame of the spectator) is lower than the potential generated 
by neighbouring spectator nucleons.  
The process of formation of spectator hyper-matter was investigate in 
Ref. \cite{botvina2011} within the transport approaches, Dubna cascade model 
(DCM) \cite{toneev83,toneev90}, and Ultra-relativistic Quantum Molecular 
Dynamics (UrQMD) model \cite{Bleicher:1999xi,Bass:1998ca}. 
It was concluded that already at beam energy of 2 A GeV hyper-spectators 
with one absorbed $\Lambda$ can be noticeably produced. While at energy of 
few tens of GeV per nucleons formation of double- and multi-strange 
hyper-spectators become feasible. 

Here we use the DCM transport approach, therefore, it should be recalled 
in more details. 
The DCM is based on the Monte-Carlo solution of a set of the 
Boltzmann-Uehling-Uhlenbeck relativistic kinetic equations with 
the collision terms, including cascade-cascade 
interactions. For particle energies below 1~GeV it is effective to 
consider only nucleons, pions and deltas. The model includes a proper 
description of pion and baryon dynamics for particle production and 
absorption processes. 
In the original version the nuclear potential is treated dynamically, i.e., 
for the initial state it is determined using the Thomas-Fermi approximation, 
but later on its depth is changed according to the number of knocked-out 
nucleons. This allows one to account for nuclear binding. 
The Pauli principle is implemented by introducing a Fermi distribution 
of nucleon momenta as well as a Pauli blocking factors for scattered 
nucleons. 
At energies higher than about 10~GeV, the Quark-Gluon String Model (QGSM) 
is used to describe elementary hadron collisions \cite{toneev90,amelin91}. 
This model is based on the 1/N$_c$ expansion of the amplitude for binary 
processes where N$_c$ is the number of quark colours. Different terms of 
the 1/N$_c$ expansion correspond 
to different diagrams which are classified according to their topological 
properties. Every diagram defines how many strings are created  in a 
hadronic collision and which quark-antiquark or quark-diquark pairs form 
these strings. The relative contributions of different diagrams can be 
estimated within Regge theory, and all QGSM parameters for hadron-hadron 
collisions were fixed from the analysis of experimental data. The 
break-up of strings via creation of quark-antiquark and 
diquark-antidiquark pairs is described by the Field-Feynman method 
using phenomenological functions for the fragmentation of quarks, antiquarks 
and diquarks into hadrons. The modified non-Markovian relativistic kinetic 
equation, having a structure close to the Boltzmann-Uehling-Uhlenbeck 
kinetic equation, but accounting for the finite formation time of newly 
created hadrons, is used for simulations of relativistic nuclear collisions. 
One should note that QGSM considers the two lowest SU(3) multiplets in 
mesonic, baryonic and antibaryonic sectors, so interactions between almost 
70 hadron species are treated on the same footing. 
The above noted two energy extremes were bridged by the QGSM extension 
downward in the beam energy \cite{amelin90}.

Within this model the absorption of $\Lambda$ hyperons by spectators is 
described in Ref.~\cite{botvina2011}. It takes place if a hyperon 
kinetic energy in the rest frame of the residual spectator is lower than the 
attractive potential energy, i.e., the hyperon potential. This potential 
is calculated by taking into account the local density of the spectator 
residues, which can be less than the normal nucleus density. In the 
calculations we follow the propagation of all particles including 
$\Lambda$-hyperons during the whole reaction time, up to about 100 fm/c, 
and take into consideration secondary rescattering/interaction processes, 
which may lead to the hyperon production, 
the hyperon absorption, and making free the absorbed hyperons.

The DCM was already successfully proved extensively in different kind of 
reactions for description of experimental data  including particle 
production \cite{toneev83}, fragmentation of spectators 
\cite{botvina1994}, and production of 
$\Lambda$ hyperons \cite{botvina2011}. Since in this work we show results 
for very high energy also, it is instructive to demonstrate how this model 
describe general yields of light particles, which can cause 
for secondary reactions leading to strangeness production in the spectator 
kinematic region. In Fig.~1 we show comparisons of DCM calculations 
(with the QGSM elementary interaction acts) with experimental data on 
rapidity distribution of pions obtained in proton interactions at 
$\sqrt s$=17.2 GeV \cite{na49} and at $\sqrt s$=200 GeV \cite{brahms}. 
One can see that the model is quite 
good in the reproduction of pion rapidities. It is important that they 
can go far beyond the projectile and target rapidities, which are 
$y_{cm}\approx ^{+}_{-}$2.9 and $y_{cm}\approx ^{+}_{-}$5.4 for these 
cases, respectively. Such comparisons give us some confidence 
that the model can be used for description of subsequent processes 
initiated by these particles, which also include hyperon production 
in reactions with spectator nucleons. Then, similar as we have found at 
lower energies 
\cite{botvina2011}, these hyperon can be captured by nuclear residues. 

\begin{figure}[tbh]
\includegraphics[width=0.6\textwidth]{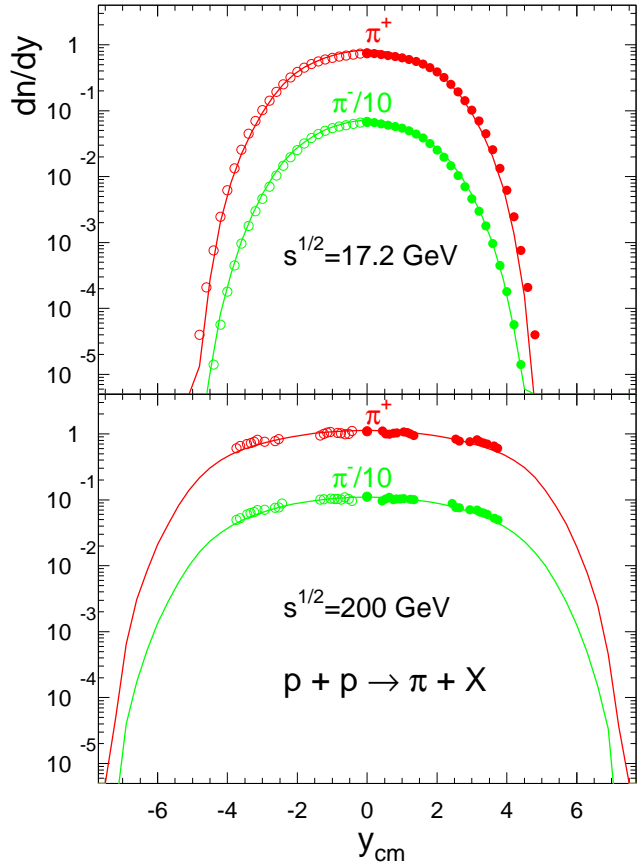}
\caption{\small{ (Color online)
Distribution of $\pi^{+}$ and $\pi^{-}$  yields versus the center-of-mass 
rapidity $y_{cm}$ in 
proton-proton collisions at energies of $s^{1/2}$=17.2 GeV (top panel) 
and $s^{1/2}$=200 GeV (bottom panel). Solid points are experimental 
data \cite{na49,brahms} 
(open ones - their symmetric reflections in the rapidity axis), 
solid lines are DCM calculations. 
}}
\label{fig1}
\end{figure}


\section{Dependence of yields of hyper-spectator residues on incident energy}

In the following we investigate how the absorption of 
$\Lambda$ hyperons by spectator residues evolves with incident projectile 
energy. As mentioned above, in the DCM calculations we use a prescription 
for the hyperon absorption elaborated in Ref.~\cite{botvina2011}, which 
gives results similar to ones obtained in other transport (UrQMD) 
calculations. 
For clarity, we consider collisions of symmetric ions, both light and 
heavy ones.

In Fig.~2 we demonstrate evolution of yields of hyper-residues in the 
very broad range of the projectile incident energies (in laboratory system) 
for carbon, nickel and lead collisions. 
The DCM calculations were done taking into account all impact parameters 
as in experiment. For example, in this case in carbon reactions more than 
90\% events contain the spectator residues with the mass number A$>$1. 
We note that usually one hyperon 
is absorbed in these reactions. The absorption of two and more hyperons is 
also possible, especially for heavy nuclei. However, probability of the 
second absorption is considerably lower \cite{botvina2011} and it does not 
influence the general behaviour of the curves. One can see a rapid 
increase of the yields with energy at low incident energies, which is 
related to the threshold character of $\Lambda$ production. The yield 
per inelastic event is much larger for the case of heavy projectile. 
This has a simple explanation: in collisions of many nucleons the 
strange particles can be more abundantly produced and more absorption 
events can take place at large residual 
nuclei. At energies around 10 GeV per nucleon we 
observe a nearly saturation behaviour. This means that we do not need 
too high energies to produce single hypernuclei. However, we should be 
careful at this point: The production of double and multiple 
hypernuclei can increase with the incident energy \cite{botvina2011}. 
This effect will be investigated in next works. 
\begin{figure}[tbh]
\includegraphics[width=0.6\textwidth]{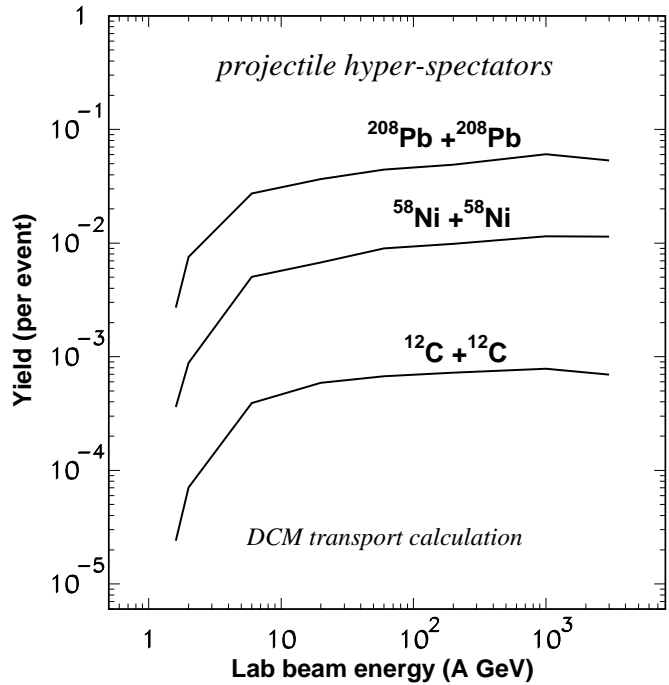}
\caption{\small{
Yields of hyper-residues of projectiles in collisions of $^{12}$C, 
$^{58}$Ni, and $^{208}$Pb beams with the same targets, as function of the 
incident energy.
The DCM calculations are integrated over all impact parameters, and 
normalized to one inelastic collision event. 
}}
\label{fig2}
\end{figure}

As was shown previously \cite{botvina2011} these hypernuclei 
residues have a broad mass distribution. Their masses are defined 
after all fast nucleons leave the residues and low-energy nucleons are 
captured inside them. The examples of such distributions at moderate 
beam energies are shown in 
Ref.~\cite{botvina2011} (for large nuclei) and in Ref.~\cite{botvina2012} 
(for small nuclei). However, as one can see from Fig.~3 the average masses 
of these residues do not change practically over all projectile energies 
under investigation. 
\begin{figure}[tbh]
\includegraphics[width=0.6\textwidth]{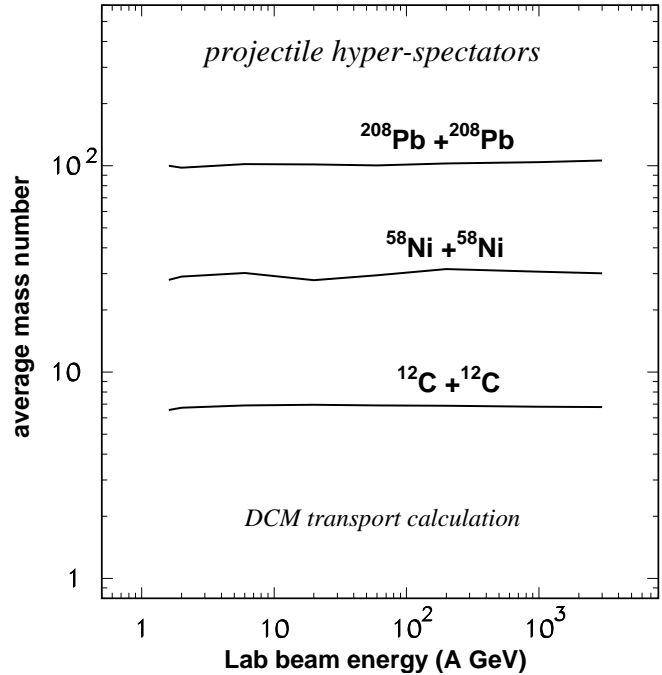}
\caption{\small{
Average mass number of the projectile hyper-residues, shown in Fig.~2, 
as function of the incident energy. 
}}
\label{fig3}
\end{figure}

It is very instructive to analyze the rapidity distribution of 
produced hyper-residues and compare it with the distribution of 
free $\Lambda$ hyperons. It is shown in Fig.~4 that such 
distributions for produced $\Lambda$s are quite wide, and evolve 
from a Gaussian-like one in central midrapidity zone at low beam energies 
to plateau-like and double-peak ones at high energy. 
This tells us that the main source 
of $\Lambda$s is not direct nucleon-nucleon collisions in the 
participant (overlapping) zone, but it is related to secondary 
hadron interactions. This point was already discussed in details in 
Ref.~\cite{botvina2011}, where the space-time evolution of the absorption 
at 20 A GeV was demonstrated, and it was shown that the DCM describes the 
rapidity distribution of $\Lambda$ well at $\sim$ 2 GeV per nucleon. 
Unfortunately, there are no experimental data for large rapidity at 
high energy. However, as we see from Fig.~1 the model is able to 
describe elementary experimental data for pion production for all 
rapidities at $\sqrt s$=17.2 corresponding to the beam energy of 158 A GeV. 
Those distributions go far beyond the projectile and target rapidities, and 
interactions of these mesons with residues contribute to the hyperon 
production too. 
\begin{figure}[tbh]
\includegraphics[width=0.6\textwidth]{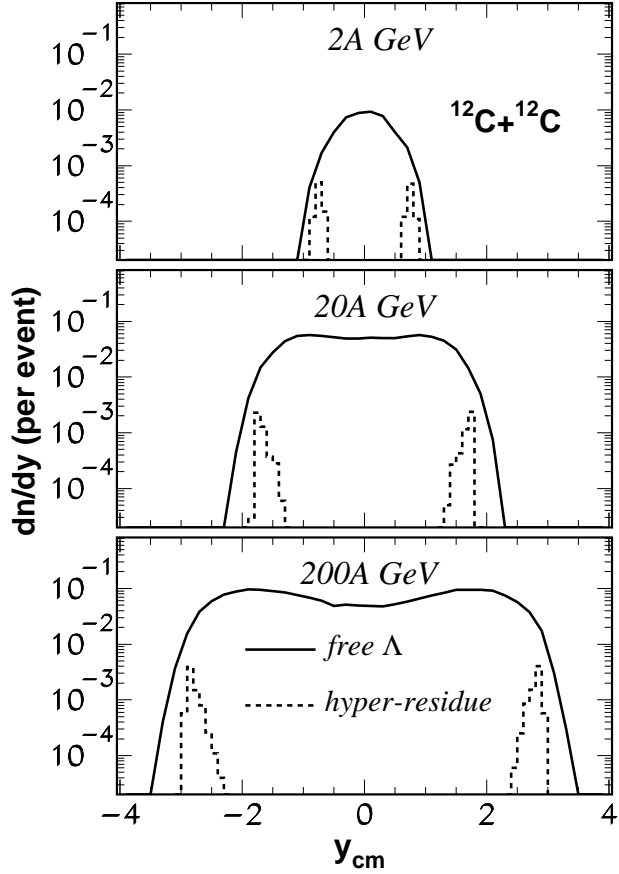}
\caption{\small{
Rapidity distributions in the system of center-of-mass ${\bf y_{cm}}$ 
of free $\Lambda$ (solid curves) and projectile and target hyper-residues 
(dashed histograms) normalized per one inelastic event in $^{12}$C+$^{12}$C 
interactions. Top, middle, and bottom panels are for collisions with 2, 20 , 
and 200 GeV per nucleon energy, respectively, as calculated with the DCM. 
}}
\label{fig4}
\end{figure}

It is natural that the existence of large nuclear residues favors both 
production and 
capture of hyperons in secondary reactions. As seen from Fig.~4, for 
rapidities higher than the projectile (and target) ones the yield of 
free $\Lambda$ drops essentially. 
It is partly related with absorption of hyperons by spectator residues. 
On the other hand decreasing the number of baryons and secondary 
particles (like pions, see Fig.~1) does not give a chance to produce many 
new $\Lambda$ hyperons in the region above the spectator rapidity.

The results shown in Figs.~2 , 3 and 4 tell us that the 
hyperon capture happens as a result of an universal stochastic process. 
As was discussed previously \cite{bot-poch,cassing,botvina2011,botvina2012} 
the production of low-energy $\Lambda$ which can be potentially captured is 
related to many rescatterings and secondary interactions involving 
produced particles. 
The yield of these hyperons is mainly determined by the 
amount of secondary particles in the vicinity of the projectile rapidity. 
However, when the energy exceeds the threshold essentially ($\goo$ 10 A GeV) 
the hyperons and other produced light particles tend to populate the rapidity 
space broadly at all beam energies: For example, one can see 
from Fig.~4 that around the residue rapidity the ratio of 'background' free 
$\Lambda$s to the hyper-residues 
remains nearly the same, a little bit less than factor 10. 
On the other hand the residues can loose certain number of nucleons 
during the particle interaction process. Depending on this interaction 
the number of hyperons can fluctuate event by event too. 
Nevertheless, there is a balance between 
the amount of hyperons with energies suitable for capture and the number of 
residual nucleons on which this capture may happen. This can 
explain that the mean mass number of hyper-residues is practically 
the same for all energies (Fig.~3). 
The predicted saturation of the production at high energies (Fig.~2) 
indicate that there are opportunities to study 
projectile- and target-like hypernuclei at different 
relativistic heavy-ion facilities (e.g., GSI/FAIR, JINR/NICA, RHIC, LHC) 
with comparable yields.

One should remember, however, that these hyper-residues are excited. 
During the next stage of the reaction they disintegrate in normal 
nuclei and nuclei containing hyperons. This process can be described 
within the models of multifragmentation \cite{bot-poch,buyuk2013} and 
the Fermi-break-up \cite{botvina2012,lorente}. The processes of fission 
and evaporation of hot hypernuclei can also take place by analogy with 
behaviour of normal excited nuclei. 
Presently, the experimental methods for identification 
of light hypernuclei are most reliable \cite{japan,saito-new}. 
For this reason and to guide future experimental studies we concentrate 
in the next sections on predictions for light ion reactions.

\section{Production of light hypernuclei}

The general information on the formation of light projectile hyper-residues 
and its evolution with beam energy is displayed in Figs.~2--4 for the case 
of carbon-carbon collisions. As was mentioned, their de-excitation caused 
by strong interaction can 
be described by the Fermi-break-up model \cite{lorente}. The procedure of 
our analysis was the following: For all inelastic $^{12}$C + $^{12}$C 
collisions we had found mass number, charge, strangeness, excitation 
energy and kinematic characteristics of the projectile residues within the 
DCM. The Fermi-break-up model (FBM) was used afterwards to describe their 
disintegration. We use the connection of dynamical and statistical models 
with parameters which have provided the best reproduction of experimental 
data analyzed in 
Ref.~\cite{botvina2012}. As was there demonstrated, reasonable variation 
of excitation energies of hyper-residues would lead to slightly different 
results. However, the results for the yields will not change by more than 
30\%, which is typical precision for the hybrid calculations. 
This is also a characteristic accuracy in description of experimental data 
within an approach including dynamical transport plus statistical decay for 
normal fragment production in such collisions of nuclei. It is sufficient 
for our present purposes intended to give a qualitative understanding of 
the processes for future experimental needs. 

It is necessary to mention that production of light clusters is sometimes 
related to coalescence of individual nucleons. However, this mechanism 
describes experimental data usually in midrapidity region, where there 
are many free particles \cite{Steinheimer,toneev83}. It looks like that 
this mechanism is responsible for production of lightest hypernuclei 
and anti-hypernuclei at relativistic central collisions at RHIC and LHC 
\cite{star,alice}. In the region of spectator residues the coalescence 
may not be effective for description of the data \cite{botvina2012}, 
and systematic comparison with experiment is needed. 

In Fig.~5 we show results of our hybrid DCM plus FBM calculations for 
carbon-carbon collisions in the projectile energy range from the 
threshold of $\Lambda$ production to 3 TeV per nucleon. The highest 
energy corresponds to LHC ion beams: In this respect we cover energies 
available with modern accelerators. We considered production of both 
well known hypernuclei (as $^{4}_{\Lambda}$H, $^{3}_{\Lambda}$H, 
$^{7}_{\Lambda}$Be) and recently discovered an exotic neutron-rich 
$^{6}_{\Lambda}$H \cite{6LH}. 
Formation of a hypothetical $\Lambda$--neutron bound state (N$\Lambda$) 
was discussed previously in Refs.~\cite{botvina2012,alice,saito}, therefore, 
we show predictions for N$\Lambda$ systems too. One can see that 
evolution of the hypernuclei's yield with energy demonstrates the same feature 
as the production of the hypernuclear residues (Fig.~2). Namely, there is a 
rapid increase around the threshold and a saturation-like behaviour at 
higher energies. It is quite obviously, since production of particular 
hypernuclei is regulated by statistical decay of excited residues, 
which properties do not change noticeably with the beam energy (see, e.g., 
masses in Fig.~3). We see that smaller hypernuclei have usually 
higher probability, mainly, because the excited hyper-residues are initially 
rather small (in average) and they disintegrate into small nuclei. 
The phase space favors also disintegration of such residues 
into small pieces, since there is no considerable energy gain coming 
from the hyperon binding in the case of formation of nuclei with larger 
mass numbers. 
\begin{figure}[tbh]
\includegraphics[width=0.6\textwidth]{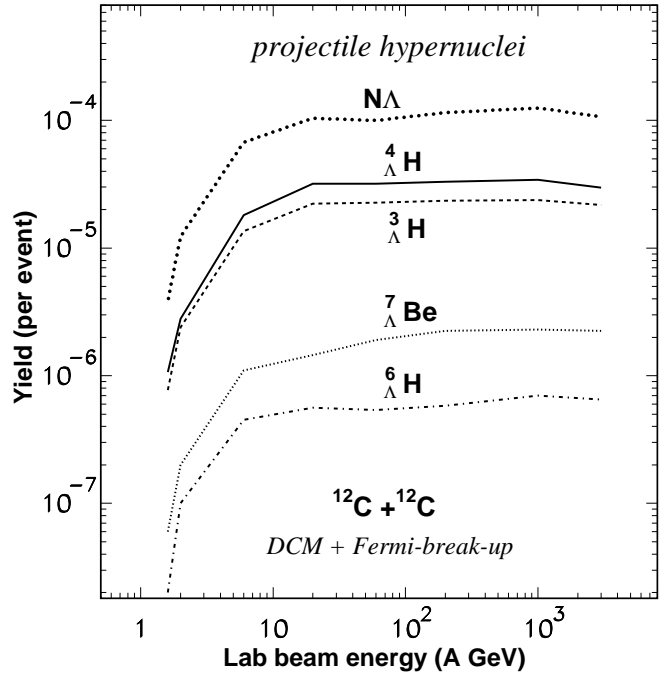}
\caption{\small{
Yields of particular hyper-nuclei (see figure and the text) obtained 
from projectile residues in collisions of $^{12}$C with $^{12}$C 
versus projectile energy in laboratory system. The hybrid DCM and FBM 
calculations are integrated over all impact parameters, and 
normalized to one inelastic collision event. 
}}
\label{fig5}
\end{figure}

It is important to analyze how the yield of particular hypernuclei 
depends on mass number and charge of colliding nuclei. This problem 
was already addressed in Ref.~\cite{buyuk2013} in relation to 
production of hypernuclei beyond neutron and proton drip-lines. In our 
case we have performed DCM and FBM calculations by taking different 
isotopes of carbon as projectiles. The beam energy of 20 A GeV was 
adopted because it is expected for FAIR facility at GSI. The considered 
beams could be easy obtained there with the FRagment Separator 
(FRS) \cite{aumann,frs}. Figure~6 demonstrates that yields of specific 
projectile hypernuclei are very sensitive to isotope composition of the 
projectile. One can considerably increase production of neutron-rich 
hypernuclei when we take neutron-rich beams (e.g., $^{16}$C). 
For example, the yield of 
exotic $^{6}_{\Lambda}$H, may increase by two orders, and this makes 
its observation much easier. If we take proton-rich beams (e.g., $^{10}$C), 
production of proton-rich hypernuclei, like $^{7}_{\Lambda}$Be, becomes 
more prominent. We believe, in reactions initiated by different isotopes 
one can obtain all kinds of hypernuclei which may exist. In addition, by 
looking at relative yields of hypernuclei we can better investigate the 
reaction mechanism and properties of hyper-matter at low temperature 
\cite{bot-poch,botvina2012}. 
\begin{figure}[tbh]
\includegraphics[width=0.6\textwidth]{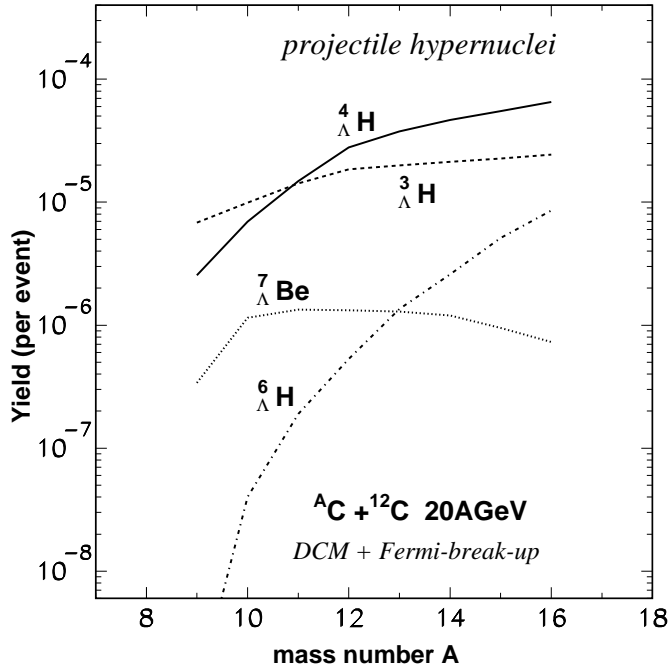}
\caption{\small{
The same as in Fig.~5, but in collisions of projectile carbon isotopes 
with target $^{12}$C versus the isotope mass number A. 
The beam energy is 20 GeV per nucleon. 
}}
\label{fig6}
\end{figure}

\section{Identification of hypernuclei}

The suggested above reaction mechanisms are very promising for 
producing hypernuclei. In particular, one can use exotic neutron-rich and 
neutron-poor projectiles \cite{buyuk2013} which are not possible to use 
as targets in traditional hyper-nuclear experiments with meson and 
electron beams \cite{Bando,japan}, because of their short lifetime. 
In addition, one can study the equation of state and phase transition in 
hypermatter at low and moderate temperatures similar as it was done in normal 
nuclear matter \cite{smm,aladin97,EOS,pochodzalla1997}. 
An essential advantage of peripheral relativistic collisions over central 
ones is that in the last ones only light hypernuclear species may be 
formed because of the high fireball temperature \cite{Steinheimer}. 
We avoid too high temperatures and we do not see any practical limitation 
on the production of large hypernuclei in peripheral collisions. 
Moreover, in such collisions we can accumulate strangeness in the 
projectile residues \cite{botvina2011} and move step-by-step to 
multi-hyperon systems. However, an important question is if it is 
possible to identify hypernuclei reliably on the background of products 
of various nuclear interactions which take place in relativistic 
nucleus-nucleus collisions. We feel rather optimistic in this respect 
because of previous successful observation of hypernuclei 
\cite{saito-new,star,alice}. However, we believe it would be very useful 
for future experiments to show model calculations of the background and 
make its comparison with the expected signal. 

Presently, the main channel for identification of light hypernuclei is 
their pionic decay. For example, a weak process  
$^{3}_{\Lambda}$H$\rightarrow \pi^{-} + ^{3}$He, with a characteristic 
lifetimes around $\sim$200 picoseconds can be seen in the 
experiments \cite{saito-new,star,alice}. The background of this process 
consists of correlations of $\pi^{-}$ and $^{3}$He coming from other 
sources: The observed $\pi^{-}$ are mainly obtained in strong interactions 
of various particles, and $^{3}$He can be produced as a result of decay of 
non-strange excited spectator residues. In Fig.~7 we show general 
rapidity distribution of $\pi^{-}$ and non-strange residues calculated 
with DCM for collisions of $^{12}$C projectile with $^{12}$C target at 
beam energies of 2, 20, and 200 GeV per nucleon. If compare it with 
a similar figure for hyperons and strange residues (Fig.~4) one can see 
that yields of non-strange particles are much larger. We see that pions 
overlap essentially the spectator residues, therefore, their correlation 
with fragments produced after disintegration of spectators must be 
investigated. 
On the other hand, we see also from Figs.~4 and 7 that for these processes 
with residues a simple 'participant--spectator' picture of reaction is not 
very precise. After nucleus-nucleus collision a residue fly in a dilute 
'cloud' of particles, mostly, light pions. Secondary interactions between 
such particles and the residue are important for the residue's strangeness 
content. 
\begin{figure}[tbh]
\includegraphics[width=0.6\textwidth]{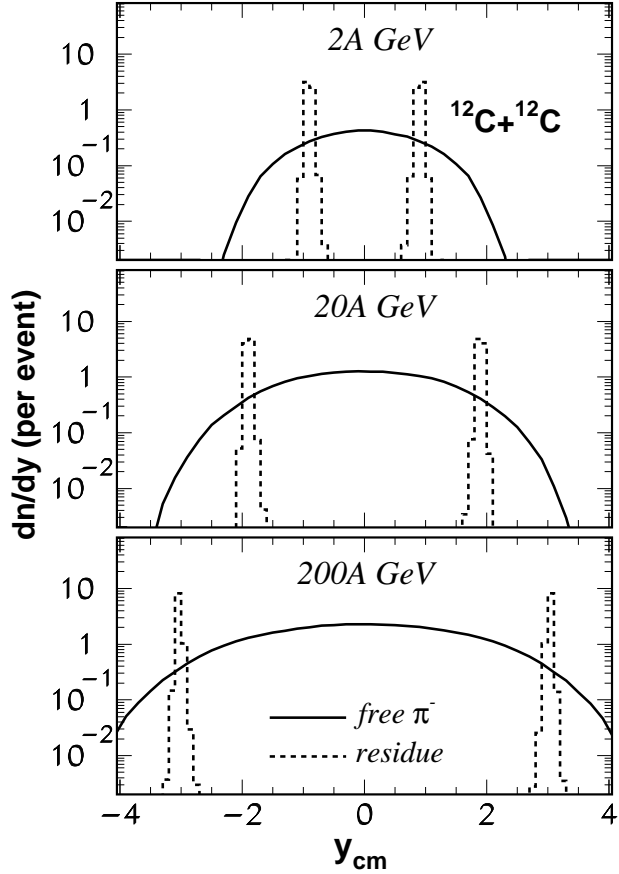}
\caption{\small{
Rapidity distributions in the system of center-of-mass ${\bf y_{cm}}$ 
of free $\pi^{-}$ (solid curves) and projectile and target spectator residues 
(dashed histograms) normalized per one inelastic event in $^{12}$C+$^{12}$C 
interactions. Top, middle, and bottom panels are for collisions with 2, 20 , 
and 200 GeV per nucleon energy, respectively, as calculated with the DCM. 
}}
\label{fig7}
\end{figure}

The corresponding DCM and FBM calculations of correlations of pions and 
fragments were performed and the results 
are demonstrated in Fig.~8: We analyze invariant masses of $\pi^{-}$ and 
deuterons, and $\pi^{-}$ and $^{3}$He. The first pair is interesting 
as decay products of a possible hypothetical $\Lambda$--neutron (N$\Lambda$) 
bound system \cite{alice,saito}. 
The second pair is typical for $^{3}_{\Lambda}$H 
weak decay. In modern experimental analyses a precision for determination 
of invariant masses is around 5--10 MeV \cite{saito-new,star,alice}. For this 
reason we have also adopted 10 MeV invariant mass intervals and counted 
event by event how many pairs fall to these intervals in the events where 
the both particles are produced. Afterwards, total normalization to the 
number of inelastic events was performed. The most interesting part of the 
distributions is a rise around the two particle threshold and its following 
transformation to a plateau-like behaviour caused by a very broad rapidity 
range of produced pions. Within $\sim$50 MeV above the threshold we expect 
a signal of the hypernucleus decay.  
\begin{figure}[tbh]
\includegraphics[width=0.6\textwidth]{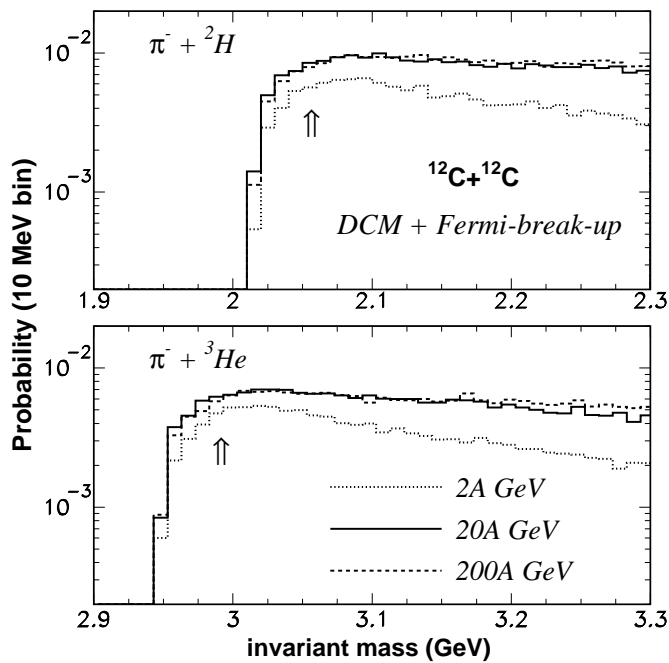}
\caption{\small{
Invariant mass distributions of the $\pi^{-}$ plus $^{2}$H pairs (top panel), 
and the $\pi^{-}$ plus $^{3}$He pairs (bottom panel) obtained in 
carbon--carbon collisions at energies of 2 GeV per nucleon (dotted 
histograms), 20 GeV per nucleon (solid histograms), and 200 GeV per nucleon 
(dashed histograms). The event by event calculations are performed within 
the DCM and Fermi-break-up models. The count number of the pairs in 10 MeV 
bins of invariant mass are normalized per inelastic event and noted as 
probability. Arrows mark the invariant masses corresponding to an 
N$\Lambda$ bound state \cite{saito} and to $^{3}_{\Lambda}$H nuclei. 
}}
\label{fig8}
\end{figure}

It is important that the signal should be clearly seen above the background. 
By examining $^{3}_{\Lambda}$H hyper-nucleus we can estimate from Fig.~5 
its probability as about 2.5$\cdot$10$^{-5}$ per event at 20A GeV. 
Whereas the background pairs in the 10 MeV interval around this 
hypernucleus mass reach $\sim$5$\cdot$10$^{-3}$ per event (Fig.~8). 
Improving energy resolution will make the signal separation better. 
However, the ratio of signal to background is expected to be around 
1\% for this nucleus. This ratio may increase to few percent for the 
N$\Lambda$ states. In these cases experimenters must take into account the 
time delay of the pionic decay and filter the corresponding events. 
The relatively long lifetime of hypernuclei allows to select a displaced 
vertex of correlated particles and to decrease 
the background significantly. For example, in central collision 
experiments, where many hundreds background pions are produced, 
it was possible to increase 
the ratio of signal to background up to 20--30\% in such a way 
\cite{star,alice}. 

The situation with identification of projectile 
hypernuclei may be better because of large $\gamma$-factors 
increasing their lifetime. In recent HypHI experiments \cite{saito-new} 
at relatively low energies (2A GeV) the signal and lifetime of projectile 
hypernuclei were reliably measured even at low statistics. These 
hypernuclei can propagate at beam energies of 2A GeV, 20A GeV, 200A GeV 
and 3A TeV about 20 cm, 130 cm, 13 meters, and nearly 200 meters, 
respectively, before their weak decay. Such a space separation 
from the place where they were produced increases chances on their 
identification, since many particles contributing to 
the background shown in Figs.~8 can be filtered out. 

Actually, the experimental set up for this measurement should be 
constructed by taking into account the accelerator energy. 
It is interesting that some already existing detectors might be 
suitable for this measurement. 
For example, at the distance around 120 meters from the ALICE detector at 
LHC a Zero Degree Calorimeter (ZDC) is located \cite{zdc}. 
Originally ZDC was designed 
to measure neutrons and one-charged particles coming from decay of 
spectator residues for better selection of central events. However, 
it is the exact place where neutral and one-charged projectile hypernuclei 
(like exotic N$\Lambda$ system and $^{6}_{\Lambda}$H) will decay. 
Many other free hyperons flying with similar rapidities will reach ZDC 
too. A fixed target providing much higher luminosity 
in comparison with colliding beams would be sufficient for this 
experiment. 


\section{Conclusion}

We investigate new promising reactions for production of hypernuclei 
in peripheral relativistic ion collisions over a wide energy range. 
Hadron interactions during dynamical stage of the process can lead to 
production of hyperons, which are captured by spectator residues. 
Disintegration of these hot residues leads to production of hypernuclei. 
This kind of reactions is well known in nuclear physics and it is 
associated with fragmentation and multifragmentation of 
relativistic residual nuclei. It is described quite reliably by 
hybrid (dynamical and statistical) models. Presently, production 
of projectile residues is widely used for synthesizing new elements, 
producing nuclei around the proton and neutron drip-lines, investigating 
the phase transition in nuclear matter at subnuclear densities. 
Including hyperons interacting in nuclear matter analogous to nucleons 
will open new opportunities for this study and can extend it in the 
direction of strangeness sector towards multi-strange nuclear systems at 
low temperatures. In addition, this method for obtaining 
hypernuclei has some advantages over the currently used ones: One can 
use exotic unstable projectiles and, as a result, to produce exotic 
hypernuclei, which may not be reachable in other ways. 

The model calculations performed in this paper indicate that this 
production mechanism is effective already at beam energies $\goo$2 GeV 
per nucleon. At the energies more than 10 GeV per nucleon a saturation 
of the yields of single hypernuclei takes place. For this reason 
these reactions can be investigated at future FAIR at GSI and 
NICA at JINR facilities, as well 
as at presently operating RHIC and LHC facilities. 
Depending on the size of colliding nuclei the cross-section for formation 
of excited hyper-residues in the saturation region is about $\sim$1--10$^2$ 
millibarn. We show also 
probabilities for production of specific light hypernuclei and evaluate the 
background conditions for their measurement. 
By selecting neutron-rich and neutron-poor isotope beams it is possible 
to increase yields of exotic hypernuclei by more than one order of 
magnitude. 
We are convinced that hypernuclear and nuclear physics will benefit strongly 
from exploring new ways for production of hypernuclei 
associated with fragmentation of spectator residues. 
 
\begin{acknowledgments}
We thank M.~Bleicher, I.~Mishustin, and T.~Saito for discussions. 
This work was supported by the Helmholtz-Institut Mainz. A.S.B. 
thanks the Frankfurt Institute for Advanced Studies J.W. Goethe University 
and the Institut f{\"u}r 
Kernphysik of J.Gutenberg-Universit{\"a}t Mainz for hospitality. 
We also acknowledge the support by the Research Infrastructure Integrating 
Activity Study of Strongly Interacting Matter HadronPhysics3 under the 7th 
Framework Programme of EU (SPHERE network). 
\end{acknowledgments}

\end{document}